\documentclass[aps,twocolumn,amsmath,amssymb,showpacs,prl,superscriptaddress,unsortedaddress]{revtex4}
\usepackage{epsf}
\usepackage{graphicx}

\begin{document}

\title{Doping-dependent energy scale of the low-energy band renormalization in (Bi,Pb)$_2$(Sr,La)$_2$CuO$_{6+\delta}$}

\author{Takeshi Kondo}
\affiliation{ISSP, University of Tokyo, Kashiwa, Chiba 277-8581, Japan}

\author{Y.~Nakashima}
\affiliation{ISSP, University of Tokyo, Kashiwa, Chiba 277-8581, Japan}

\author{W.~Malaeb} 
\affiliation{ISSP, University of Tokyo, Kashiwa, Chiba 277-8581, Japan}

\author{Y.~Ishida} 
\affiliation{ISSP, University of Tokyo, Kashiwa, Chiba 277-8581, Japan}

\author{Y.~Hamaya}
\affiliation{Department of Crystalline Materials Science, Nagoya University, Nagoya
464-8603, Japan}

\author{Tsunehiro~Takeuchi}
\affiliation{Department of Crystalline Materials Science, Nagoya University, Nagoya
464-8603, Japan}
\affiliation{EcoTopia Science Institute, Nagoya University, Nagoya 464-8603, Japan}

\author{S.~Shin} 
\affiliation{ISSP, University of Tokyo, Kashiwa, Chiba 277-8581, Japan}
\affiliation{CREST, Japan Science and Technology Agency, Tokyo 102-0075, Japan}

\date{\today}
\begin{abstract}
{The nodal band-dispersion in  (Bi,Pb)$_2$(Sr,La)$_2$CuO$_{6+\delta}$ (Bi2201) 
 is investigated over a wide range of doping by using 7-eV laser-based angle-resolved photoemission spectroscopy.
We find that the low-energy band renormalization ($``$kink"), recently discovered in Bi$_2$Sr$_2$CaCu$_2$O$_{8+\delta}$ (Bi2212),
also occurs in Bi2201, but at a binding energy around half that in Bi2212, implying its scaling to $T_{\rm c}$.   Surprisingly the coupling-energy dramatically 
increases with a decrease of carrier concentration, showing a sharp enhancement across the optimal doping. 
This strongly contrasts to  other mode-couplings at higher binding-energies ($\sim$20, $\sim$40, and $\sim$70 meV) with almost no doping variation in energy scale. 
These nontrivial properties of the low-energy kink (material- and doping-dependence of the coupling-energy) 
demonstrate the significant correlation among the mode-coupling, the $T_{\rm c}$, and the strong electron correlation.
}
\end{abstract}


\maketitle
The band renormalization or $``$kink" in cuprates have been attracting huge interest in the condensed matter physics because of a prospect that the 
associated collective mode plays an essential role in the pairing of electrons causing the high temperature superconductivity. However, the nature of the kink, particularly whether it is due to phonon or spin excitations, 
remains controversial mostly because the different modes occur at almost the same energy \cite{OldKink}.  
Previously it was reported that the nodal group velocity within 50 meV of Fermi energy ($E_{\rm F}$) is independent of the cuprate family or the number of CuO$_2$ layers in the crystal unit cell, and it is almost constant across the entire phase diagram \cite{Zhou_Nature}. Recent ARPES technique with low-energy photons ($h\nu$ = 6-8 eV) has extensively improved the momentum and energy resolutions\cite{Kiss,Ishizaka_Laser}, and uncovered a remarkable band renormalization very close to the $E_{\rm F}$ ($<$ 20 meV) \cite{Zhou_Laser,Johnson _Laser,Shen_Laser,Anzai,Dessau_Laser,Johnston_Laser}, in addition to the well-studied kinks seen at 40-80 meV, in Bi$_2$Sr$_2$CaCu$_2$O$_{8+\delta}$ (Bi2212) [${T_{\rm{c}}^{\max }}$ = 95 K].
The new fine band-feature (or $``$low-energy kink") determines the nodal Fermi velocity of Bi2212, thus it is crucial for the understanding of the electronic properties, which are dominated by the conduction electrons close to the $E_{\rm F}$.

A recent theoretical study has suggested that coupling to phonons is too small to produce the observed band renormalization \cite{Giustino_Nature}.
On the other hand,  it has been pointed out that the strong electron correlation or reduced screening can significantly enhance the phonon-electron coupling \cite{Devereaux_Nature,correlation1,correlation2,correlation3,Johnston_Laser}. 
A systematic study of the fine band-feature from the metallic overdoped-region to the poorly screened underdoped-region is crucial to 
reveal the mechanism of the mode-couplings in cuprates. 
To address the issue, we chose  (Bi,Pb)$_2$(Sr,La)$_2$CuO$_{6+\delta}$ (Bi2201) [${T_{\rm{c}}^{\max }}$ = 35 K] for a study,
where a wide doping range from the underdoped- to the heavily overdoped-region up to outside the $T_{\rm c}$-dome is accessible.
The studying of Bi2201 with a  $T_{\rm c}$, which is $\sim$2.5 times lower than that of Bi2212,
 is also important to confirm the universality of the low-energy kink in cuprates.

In this letter, we find low-energy kink in Bi2201 ($<$10 meV) similar to that reported for Bi2212 \cite{Zhou_Laser,Johnson _Laser,Shen_Laser,Anzai,Dessau_Laser,Johnston_Laser},  by means of ARPES with 7 eV laser. 
The energy scale of the coupling, however, is almost half of that in Bi2212, implying 
the correlation between the low-energy renormalization and $T_{\rm c}$ or the size of the energy gap.
The doping variation of the coupling is rather dramatic:
toward underdoping, the energy scale and the coupling constant monotonically increase with an abrupt enhancement across the optimal doping.
This suggests that the electron correlation is a key factor for the development  of the low-energy renormalization.
We also uncover multiple mode-couplings at higher binding energies of $\sim$20 and $\sim$40 meV in addition to $\sim$70 meV, and confirm that 
the variation of energy scale with doping is unique in the low-energy coupling, indicating that it has a different origin. 
The nontrivial behavior along the node found in this work supports the theoretical idea suggesting the  forward scattering arising from the interplay between the electrons and in-plane polarized acoustic phonon branch as the origin of the low-energy renormalization.

Single crystals of  (Bi,Pb)$_2$(Sr,La)$_2$CuO$_{6+\delta}$ (Bi2201) and Bi$_2$Sr$_2$CaCu$_2$O$_{8+\delta}$  (Bi2212) were grown by the floating-zone (FZ) technique.  ARPES measurements were performed using a Scienta R4000 hemispherical analyzer with a ultraviolet laser ($h\nu$ = 6.994eV) at The Institute for Solid State Physics (ISSP), The University of Tokyo \cite{Kiss}. The energy resolution was less than 1 meV. The samples were cleaved $in\ situ$, and kept under  a vacuum better than $3 \times {10^{ - 11}}$ torr during the experiments.

\begin{figure}
\includegraphics[width=3.5in]{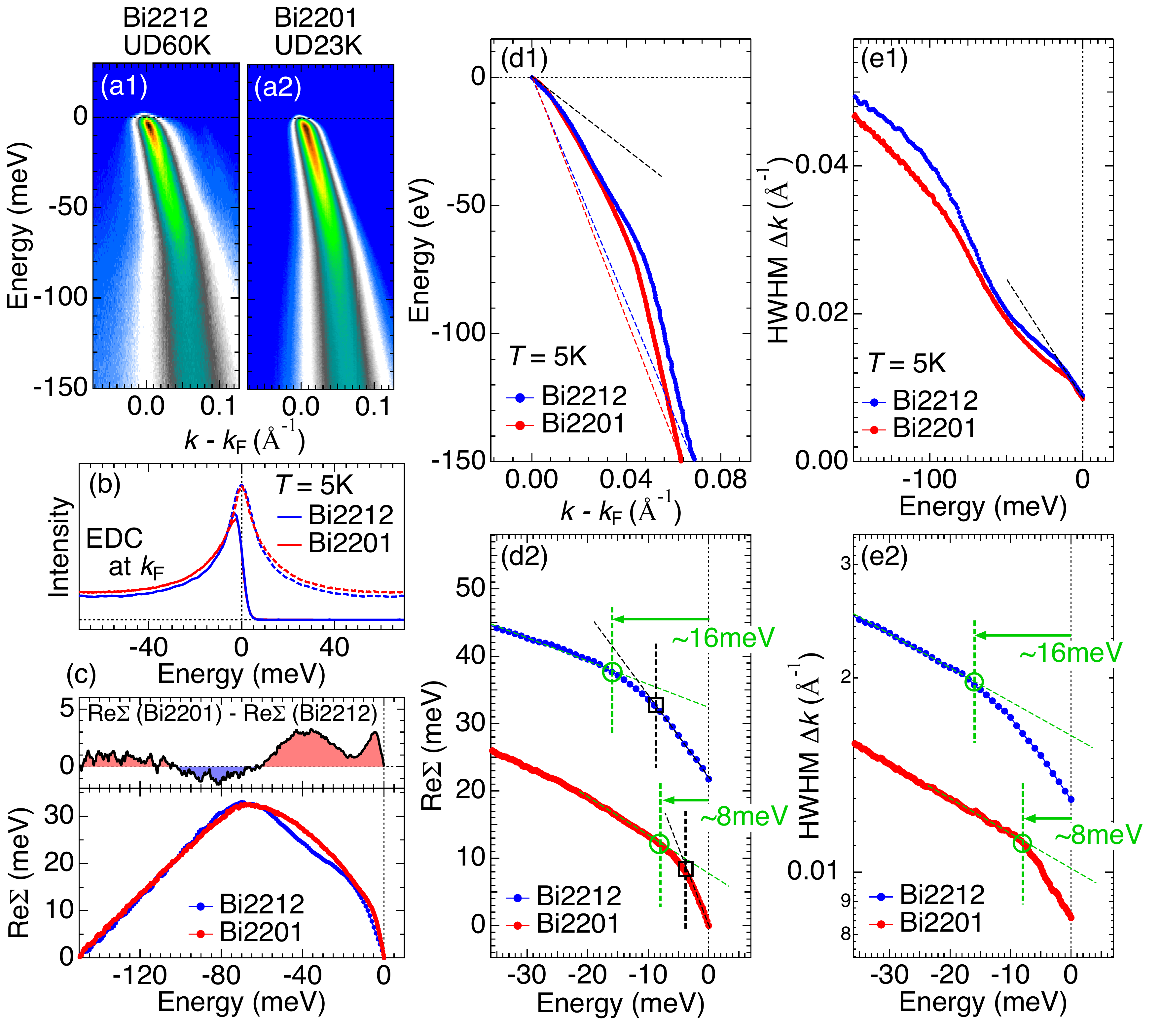}
\caption{(Color online)
Comparison of the nodal data between Bi2212 ($T_{\rm{c}}$/$T_{\rm{c}}^{\max}$ = 0.63) and Bi2201 ($T_{\rm{c}}$/$T_{\rm{c}}^{\max}$ = 0.66).  (a1)-(a2) ARPES image. (b) EDCs at $k_{\rm F}$ and the corresponding symmetrized EDCs.  (c) Re$\Sigma(\omega)$ obtained from the data in (d1).  The upper panel plots the difference spectrum of Re$\Sigma$ between the two samples.
(d1) MDC-derived band dispersion. (e1) Peak widths of MDCs. 
(d2) Same data as in (d1) magnified close to $E_{\rm F}$. An offset is used for clarity. Green circle and dashed line (Black square and dashed line) indicates deviation from the behavior at the lower (higher) energies. 
(e2) Same data as in (e1) magnified close to $E_{\rm F}$.  The data is shown with an offset and at the logarithmic scale.  
 } 
\label{fig1}
\end{figure}

\begin{figure*}
\includegraphics[width=6.5in]{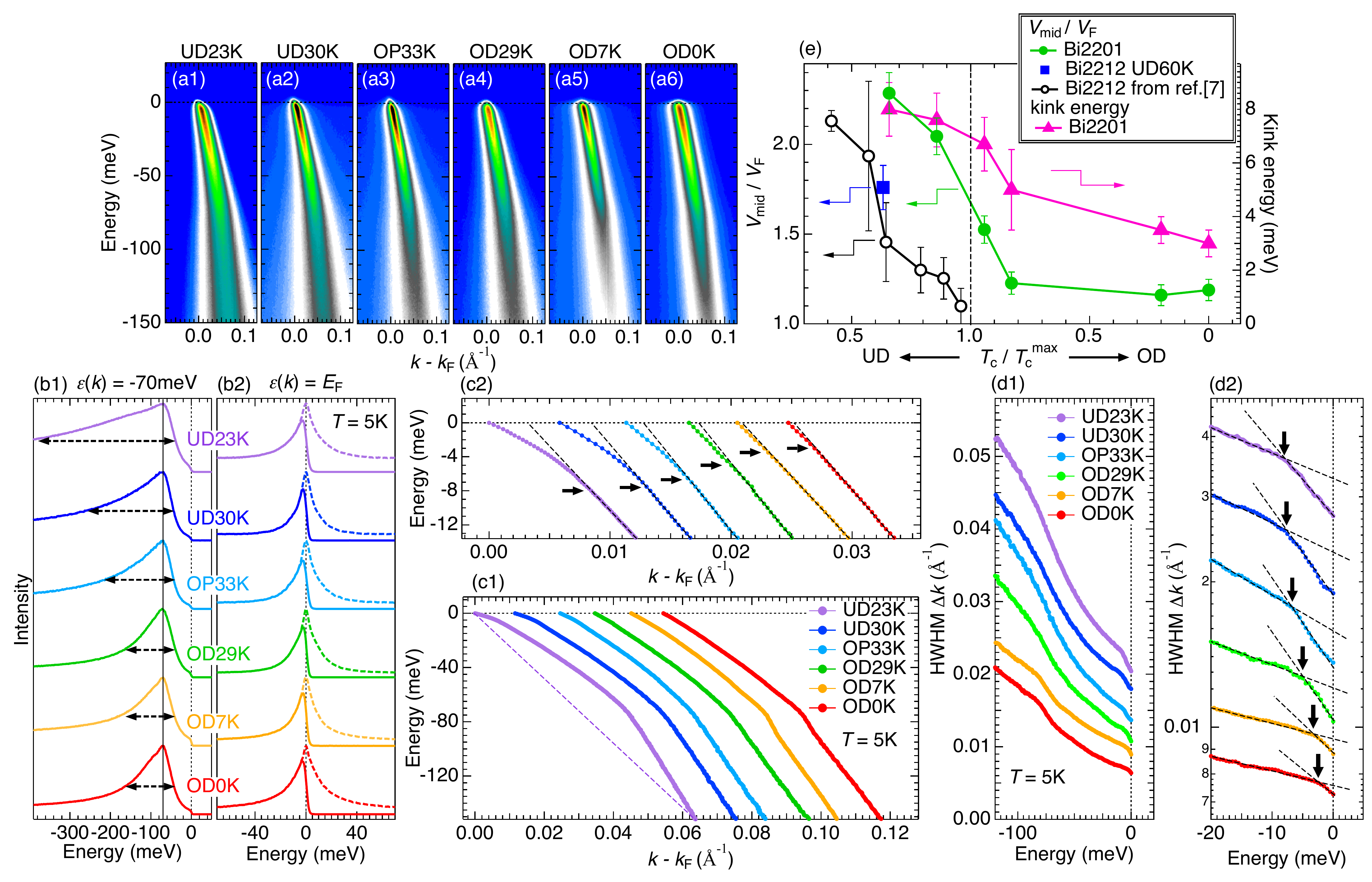}
\caption{(Color online)
Doping dependence of nodal data for Bi2201. (a1)-(a6) ARPES image. (b1)-(b2) EDCs at $\varepsilon (k) = -70$ meV and $E_{\rm F}$, respectively. Arrows in (b1)
indicate the spectral widths.
Dashed curves in (b2) are the symmetrized EDCs. (c1) MDC-derived band dispersions. (c2) Same data as in (c1) magnified close to $E_{\rm F}$.  (d1) MDC peak-width plotted with an offset. (d2) Same data as in (d1) magnified close to $E_{\rm F}$ at the logarithmic scale. The arrows in (c2) and (d2) indicate energy positions of the low-energy kink. (e) The estimated kink energy and the velocity change across the kink ($V_{\rm mid}/V_{\rm F}$).}  
\label{fig1}
\end{figure*}

In Fig.1, we compare the nodal band dispersion of underdoped Bi2212 with $T_{\rm{c}}$ = 60 K (UD60K) and Bi2201 with
$T_{\rm{c}}$ = 23 K
(UD23K),  measured at the same experimental condition. 
For a fair comparison, we used samples having almost the same 
ratio between $T_{\rm{c}}$  and the optimal $T_{\rm{c}}$ for each cuprate family (${{{T_{\rm{c}}}} \mathord{\left/
 {\vphantom {{{T_{\rm{c}}}} {T_{\rm{c}}^{\max }}}} \right.
 \kern-\nulldelimiterspace} {T_{\rm{c}}^{\max }}}$ = 0.63 and 
${{{T_{\rm{c}}}} \mathord{\left/
 {\vphantom {{{T_{\rm{c}}}} {T_{\rm{c}}^{\max }}}} \right.
 \kern-\nulldelimiterspace} {T_{\rm{c}}^{\max }}}$ = 0.66 for Bi2212 and Bi2201, respectively).  
 Figures 1(a1) and 1(a2) show  the ARPES images  for Bi2212 and Bi2201, respectively. 
 The energy distribution curves (EDCs) at $k_F$ extracted in  Fig.1(b) have almost identical shape with a  sharp peak, which ensures the relevance of comparison between the two samples.  We have determined the band dispersion from the peak positions of momentum distribution curves (MDCs), and plotted it in Fig. 1 (d1). 
The low-energy kink is seen in Bi2212, as reported elsewhere \cite{Zhou_Laser,Johnson _Laser,Shen_Laser,Anzai,Dessau_Laser,Johnston_Laser}.  
Our new finding is that a similar fine-feature is observed in Bi2201, indicating that 
the  low-energy mode-coupling is an universal property of cuprates. 
In  Fig.1 (e1), we plot the peak width of MDC ($\Delta k$), which is proportional to 
${\mathop{\rm Im}\nolimits} \Sigma$ ($\Sigma$: self-energy).
A low-energy kink is clearly seen in the $\Delta k(\omega )$, as expected from the Kramers-Kronig relation between ${\mathop{\rm Re}\nolimits} \Sigma$  and  ${\mathop{\rm Im}\nolimits} \Sigma$.  We extract ${\mathop{\rm Re}\nolimits} \Sigma$ in Fig.1(c) and examin it close to $E_{\rm F}$ in Fig.1(d2). 
The ${\mathop{\rm Re}\nolimits} \Sigma(\omega )$ has been obtained from 
the energy difference between the measured dispersion and 
 the linear one [dashed lines in Fig.1(d1)] expected when the mode-couplings are absent.
In Fig.1(e2), we show the data of Fig.1(e1) magnified near $E_{\rm F}$.  For clarity, an offset is used both in Figs.1(d2) and 1(e2). 
Surprisingly the energy scale of the kink in Bi2201 ($\sim$8 meV) is almost half of that in Bi2212 ($\sim$16 meV).
Note that, while the magnitude of the kink energy depends on the estimation criteria as demonstrated in Fig.1(d2) [see green circles and black squares], the difference by a factor of 2 between the two compounds is robust regardless of the estimation schemes.
This result  implies a correlation between the low-energy mode-coupling and the $T_{\rm c}$ of the compounds.
We have also estimated the coupling constant, $\lambda  =  - {[\partial ({\mathop{\rm Re}\nolimits} \Sigma )/\partial \omega ]_{\omega  = 0}}$, and obtained  a larger value in Bi2201 ($\lambda$ = 2.32) than in Bi2212 ($\lambda$ = 1.26) by a factor of $\sim$2. This means that the coupling is much stronger in the former.

To reveal the cause of change in the coupling energy and strength,  
we investigated the doping dependence of the nodal dispersion in Bi2201.
 Figures 2 (a1)-(a6) show the ARPES maps along the node for samples with various carrier concentrations from 
 the underdoped-region [UD23K in Fig.2 (a1)] to  the heavily overdoped-region outside  $T_{\rm c}$-dome [OD0K in Fig.2 (a6)].
 The MDC-derived dispersions and the MDC peak widths ($\Delta k$s) of these samples are plotted in  Figs. 2(c1) and 2(d1), respectively. 
All of the EDCs at $k_{\rm F}$, plotted in Fig.2(b2), have sharp peaks at $E_{\rm F}$  with weak doping dependence. This allows us to do a fair comparison among the samples. 
In Fig. 2 (c2), we examine the dispersions close to  $E_{\rm F}$. 
It is clearly seen that the kink structure becomes more pronounced with underdoping. 
This behavior has been reported for  Bi2212 \cite{Shen_Laser}. 
The new result we found is that 
the coupling-energy, marked with arrows, monotonically increases toward underdoping. 
In Fig.2(e), we summarize the doping evolution of the kink energy and 
 the velocity change across the kink, $V_{\rm mid}/V_{\rm F}$ ($V_{\rm F}$: Fermi velocity, $V_{\rm mid}$: velocity just below the kink energy).
 In both of the plots, monotonic increase toward underdoping with an abrupt enhancement across the optimal doping is clearly seen. In the same panel, we also plot both our result and the published result \cite{Shen_Laser} of $V_{\rm mid}/V_{\rm F}$ for  Bi2212.  As mentioned above, the coupling is stronger in Bi2201 than in Bi2212 along the whole doping levels. The coupling in Bi2201 extends up to the heavily overdoped-region, while it seems negligible in Bi2212. This would be a consequence of the stronger coupling in Bi2201.

The uniqueness of the low-energy kink gets clearer by comparing it with mode-couplings at higher binding energies. 
In Fig.3(c1), we investigate the doping dependence of  the ${\mathop{\rm Re}\nolimits} \Sigma (\omega )$.
The values of ${\mathop{\rm Re}\nolimits} \Sigma $ were estimated from the energy difference between the measured dispersion and 
the linear one as drown in Fig.2(c1) for UD23K.
The area underneath the ${\mathop{\rm Re}\nolimits} \Sigma$ curves, as an indication of the total coupling strength, monotonically increases with underdoping.
To examine the doping variation in more detail, we normalized all the curves to each area, and plotted those in Figs. 3(a2) and 3(a3) without and with an offset, respectively.
While the overall shape is almost the same for all the samples, small but clear difference is seen around -70 meV and very close to $E_{\rm F}$. 
We find two pronounced kinks at $\sim$-40 and $\sim$-20 meV, in addition to ones around -70 meV and -5 meV. 
 These are marked with bars in Fig.3(a3). These kinks are more clearly demonstrated in Figs. 3(b1), 3(b2), and 3(b3), which show 
 magnified  Fig.3(a1)  within small energy windows [arrows in Fig.3(a3)].
 Here we note that the multiple-kinks at $ - 60 \le \omega  \le  - 20$ meV have not been observed along the node in Bi2212.
 In contrast, there are several reports  presenting such features for the low $T_{\rm c}$ materials such as  (La$_{2-x}$Sr$_x$)CuO$_4$ \cite{Zhou_PRL} as well as Bi2201 \cite{Meevasana,Zhou_Bi2201}.  We can confirm this situation in  Fig.1(c), where the ${\mathop{\rm Re}\nolimits} \Sigma (\omega )$ of Bi2201 and Bi2212 are compared.
 While the areas of the two curves are almost the same,  
a significant difference is seen in the energy range of $- 60 \le \omega  \le  - 20$ meV in addition to around -5 meV.  
This indicates that these multiple-couplings are characteristic of the low $T_{\rm c}$ cuprates, and would be 
 related with the shape of their Fermi surfaces having less parallel segments near the zone edge \cite{Ari,Devereaux_PRL}. 
 We find that the couplings  at $\sim$-40 and $\sim$-20 meV do not show clear doping dependence in energy scale, similarly to that at $\sim$-70 meV.
 This strongly contrasts to the significantly doping-dependent behavior seen in the low-energy kink, which indicates that it has a different origin.   
In Fig.3(c), we plot the second derivative of ${\mathop{\rm Re}\nolimits} \Sigma (\omega )$ [$ - {{{\partial ^2}({\mathop{\rm Re}\nolimits} \Sigma )} \mathord{\left/
 {\vphantom {{{\partial ^2}({\mathop{\rm Re}\nolimits} \Sigma )} \partial }} \right.
 \kern-\nulldelimiterspace} \partial }{\omega ^2}$], which is sensitive to the sharpness of the mode-coupling.  We found that the peak at $\sim$-70 meV is strongly suppressed  in the underdoped-region. The energy broadening of the coupling is attributed to the increase of the electron correlation. The effect is observed in Fig.2(b1), where an abrupt broadening of EDC at the energy state of $\varepsilon (k)$ = -70 meV  is seen in the underdoped-region. 
In contrast to that at $\omega\sim -70$ meV, the peak at $\omega\sim -5$ meV in   
$ - {{{\partial ^2}({\mathop{\rm Re}\nolimits} \Sigma )} \mathord{\left/
 {\vphantom {{{\partial ^2}({\mathop{\rm Re}\nolimits} \Sigma )} \partial }} \right.
 \kern-\nulldelimiterspace} \partial }{\omega ^2}$
  [Fig.3(c)]  significantly increases in the underdoped-region. 
  The anticorrelation between the couplings of the two different energies is demonstrated in Fig.3(d).   
  In the same panel, we also plot the peak width of EDC ($\Delta \varepsilon $) at $\varepsilon (k)$ = -70 meV.
 The increase with underdoping of the electron scattering and the low-energy coupling both occur at almost the same doping level, indicating that the two phenomena are tied with each other.

\begin{figure}
\includegraphics[width=3.45in]{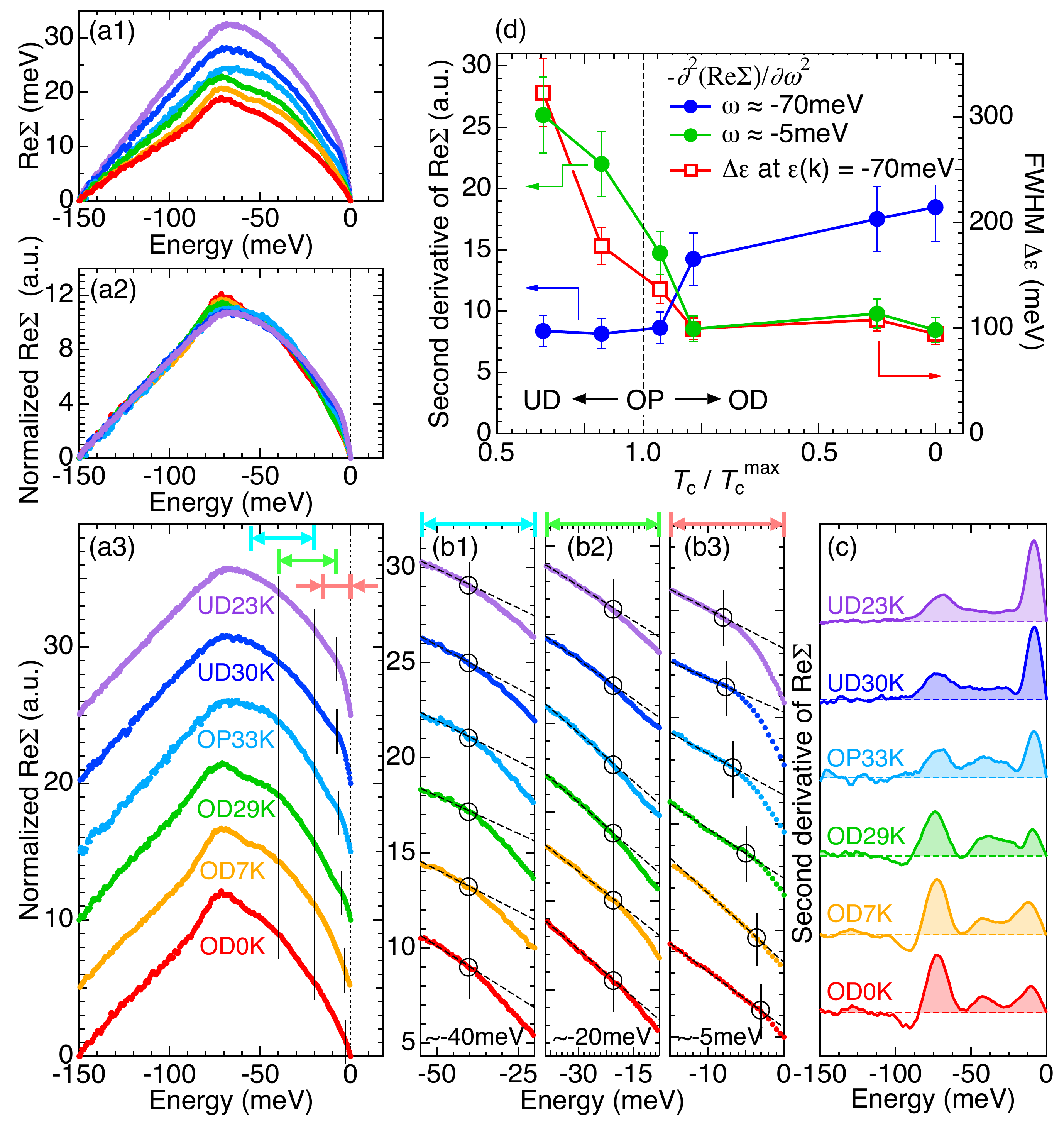}
\caption{(Color online)
(a1) Re$\Sigma$ extracted from the data in Fig.2(c1).  (a2)-(a3) Same data as in (a1) normalized to the area underneath each curve, plotted  without and with offset, respectively. The bars in (a3) indicate energy positions of the kinks. (b1)-(b3) Magnified (a3) within a narrow energy window [arrows in (a3)]. The kink position is indicated with circles and bars. 
 (c) Second derivative of Re$\Sigma$ in (a3), $ - {\partial^2}({\mathop{\rm Re}\nolimits} \Sigma )/\partial{\omega ^2}$. }  
\label{fig1}
\end{figure}
 
 A recent theoretical work \cite{Johnston_Laser} have reproduced the low-energy kink of Bi2212  in the context of coupling to the in-plane polarized acoustic phonon branch with a poor screening realized in the underdoped-region.  
 When the metallicity breaks down in the underdoped-region, 
  the coupling between the electrons and acoustic phonons, which arises via the modulation of the screened Coulomb potential, 
results in the forward scattering from momentum state $k$ to $k+q$ with small $q$.  
The nodal self-energy is, therefore,  determined by scattering to nearby states with a small superconducting gap close to the node.
This scenario can generate a kink even at a binding energy smaller than the magnitude of the antinodal gap (${\Delta _0} \approx 15$ meV in the optimally doped Bi2201). Furthermore, it  allows the variation of the coupling energy for the identical phonon branch, depending on the degree of the electron correlation (or poor screening of the conduction electrons) and the magnitude of the energy gap.
Our results showing significant doping- and material-dependence in the low-energy kink support this scenario.
Scanning tunneling spectroscopy (STM) studies have uncovered the intrinsic inhomogeneous properties in the local density of state \cite{Pan}, and 
revealed a direct relationship of it with the poor metallic nature in the underdoped-region \cite{McElroy}.   
 The stronger coupling observed in Bi2201 than in Bi2212 [see Fig.2(e)] might be related with the STM results that the inhomogeneous character is more pronounced in the former \cite{Eric,Mashima1,Mashima2}.
Interestingly, the inhomogeneous state has been observed even in the heavily overdoped samples outside the $T_{\rm c}$-dome in Bi2201 \cite{Mashima2}. This is compatible with  the existence of the low-energy kink in the same samples. Finally, it would be worth to note that a recent inelastic x-ray scattering experiment for Bi2201 has observed an anomalous broadening of the acoustic longitudinal mode at small $q$ vector, indicative of a strong coupling with electrons\cite{X-ray}. It is, however, claimed that the anomaly is suppressed in the Pb-doped samples, which are used  in the current work, for no clear reason.  Further studies would be required to address the cause and nail down the relation with our results.

In conclusion, we found the low-energy kink in the nodal dispersion of Bi2201. The energy scale is almost the half of that in Bi2212, indicating a correlation between the coupling and the $T_{\rm c}$. Our systematic study of doping variation for Bi2201 revealed the crucial role of the electron correlation on the kink structure. 
This is compatible with the theoretical idea suggesting the  forward scattering arising from the interplay between the electrons and in-plane polarized acoustic phonon branch as the origin of the low-energy renormalization.
 
This research is supported by JSPS through its FIRST Program.

\end{document}